\documentclass[preprint,tightenlines,eqsecnum,floats,aps,amsmath,amssymb,nofootinbib,prd,showpacs]{revtex4}

\usepackage{amssymb}
\usepackage{stmaryrd}
\usepackage{amsmath}
\usepackage{amsfonts}
\usepackage{mathrsfs}
\usepackage{CJK}
\usepackage{amsmath,amssymb,amsfonts}
\usepackage{graphicx}
\usepackage{subfigure}

\def\be{\begin{equation}}
\def\ee{\end{equation}}
\def\ba{\begin{eqnarray}}
\def\ea{\end{eqnarray}}

\begin{document}

\title{Collisional Penrose process of BTZ black holes}

\author{Xianglong  Wu}
\affiliation{Department of Physics, South China University of Technology, Guangzhou 510641, China}

\author{ Xiangdong Zhang\footnote{Corresponding author. scxdzhang@scut.edu.cn}}
\affiliation{Department of Physics, South China University of Technology, Guangzhou 510641, China}

\date{\today}


\begin{abstract}

	The Penrose process in the vicinity of an extremal Ban\~ados-Teitelboim-Zanelli(BTZ) black hole is studied. Due to the existence of negative cosmological constant, only massless particles could escape to infinity. Hence we analyse the Penrose process by one massless particle collides with another massive particle near the horizon of BTZ black holes. Calculations of the maximum energy extraction efficiency of this process is carried out for both spinless and spinning particles. Our results show that the spinning particle have a higher energy extraction efficiency than the spinless particle. Moreover, our calculation also indicates that the maximum energy extraction efficiency is independent of the value of the cosmological constant of BTZ black holes.
\end{abstract}
\maketitle
\section{INTRODUCTION}

It is well known that the rotational energy can be extracted from a Kerr black
hole, which is usually referred as the Penrose process. This remarkable idea was first proposed by R. Penrose \cite{Gravitational collapse: the role of general relativity}. Note that the particle can have
negative energy in the ergosphere region. Hence, if a
composite particle with energy $E_{in}$ falling into the ergosphere region and breaks up into two
particles such that one particle has negative energy and falls
into the rotating black hole, while the other particle with positive
energy $E_{out}$, which is larger than the input energy $E_{in}$, escape to
infinity. With such process, we can successfully extract energy from a rotating
black hole. The maximal efficiency of the energy extraction of the Kerr black hole was calculated by Wald in \cite{Energy limits on the penrose process}. Later, this result was extended to the Kerr-Newman black holes \cite{Energetics of the Kerr-Newman black hole by the Penrese process}. Along this line, Piran et al. investigate the collisional Penrose process \cite{High efficiency of the Penrose mechanism for particle collisions}.

On the other hand, Ba\~nados, Silk and West(BSW) shows that a rotating black hole can act as accelerators for spinless particles\cite{PhysRevLett.103.111102} in 2009. Inspired by their work, some authors suggest to consider the collision Penrose process through the BSW mechanism\cite{PhysRevLett.109.121101,PhysRevD.86.024027,PhysRevLett.113.261102}. These collision process are usually has far more higher energy contraction efficiency. For instance, the result given by Schnittman et. al. in \cite{PhysRevLett.113.261102} indicates that the maximal efficiency is about 13.92 for a massless and a massive particle collide near the horizon. Many effects have been done to generalize the collisonal Penrose process to various black holes background, see \cite{PhysRevD.82.083004,PhysRevD.82.103005,PhysRevD.83.084041,PhysRevD.83.044013,PhysRevD.93.084025,PhysRevD.97.024003,PhysRevD.99.064022}
and references therein for detail.

The BSW mechanism has been generalized to the spinning particles situation recently\cite{Armaza2015Can}. In spinning case, the orbit of the test particle deviates from geodesic\cite{A.Papapetrou.Spinning_test-particles,Dixon1970Dynamics1,Dixon1970Dynamics2,Wald1972Gravitational,Kerr1963Gravitational}. The equations of motion thus is quite different with the spinless situation. The corresponding collision Penrose process also have been investigated in
many cases\cite{PhysRevD.97.064024, PhysRevD.98.044006, PhysRevD.98.064027, Okabayashi:2019wjs}.

In (2+1)-dimensional spacetime, contrast with the (3+1)-dimensional case, there was no asymptotic flat black hole solutions for Einstein gravity. However, it was remarkable that Ba\~nados-Teitelboim-Zanelli (BTZ) found in 1992 \cite{Phys. Rev. Lett.1849}, the stationary and rotating black hole solution exist when we inclusion a negative cosmological constant. The BTZ black holes, because of its similarity and simplicity compared with the (3+1)-dimensional Kerr black hole, has received increasing attentions. In particular, the particles collision with spinless case as well as spinning case around
BTZ black hole has been invsetigated in \cite{tsukamoto2017,yuanzhang19}. Inspired and motivated by these results, in this paper, we study the collisional Penrose process around the BTZ black holes.

This paper is organized as follows: We present an introduction in Section I. Then in Section II, by using the Mathission-Papapetrou-Dixon(MPD) equations, we calculate the equations of motion of particles in the BTZ spacetime, and some constraint on the orbits are also given. In Section III, we study the collisional Penrose process of one massless particle with one massive particle (which is referred as Compton scattering), and calculate the maximum energy extraction efficiency of this process. The conclusion is given in Section IV. Throughout the paper, we adopt the geometrical units $G=c=1$.

\section{Equations of motion of spinning particles in BTZ spacetime}

\subsection{Equations of motion of a spinning particle}
We consider a spinning particle in a three-dimensional BTZ black hole spacetime. It's motion can be dictated by the Mathission-Papapetrou-Dixon (MPD) equations\cite{A.Papapetrou.Spinning_test-particles,Dixon1970Dynamics1,Dixon1970Dynamics2} as
\begin{eqnarray}
\label{equation one}
	\frac{Dp^a}{D \tau}&=&-\frac{1}{2} R^{a}{}_{bcd} v^{b}S^{cd},\\
	\label{equation two}
	\frac{D S^{ab}}{D \tau}&=&p^a v ^b-p^b v^a,
\end{eqnarray}
where $v^a = (\frac{\partial}{\partial\tau})^a$, $p^a$, and $S^{ab}$ are the velocity, momentum and antisymmetric spin tensor of particles, respectively.

The relation between the antisymmetric spin tensor $S^{ab}$ and the particle's spin $s$ is
\begin{eqnarray}
\frac{1}{2}S^{ab} S_{ab}&=&m^2 s^2,
\end{eqnarray}
where $m$ is the mass of the particle. Moreover, the momentum satisfy the normalization condition as
\begin{eqnarray}
p^a p_a&=&-m^2.
\end{eqnarray}
We introduce a special momentum $u^a$ for massive particles by
$u^a$=${p^a}/{m}$. To determine the motions of the spinning particles, we need to add the following two supplementary equations\cite{PhysRevD.93.084025, Quantum  Grav. 33:105014}:
\begin{eqnarray}
\label{equation  three}
p^a v_a&=&-m,\\
\label{equation  four}
S^{ab} p_b&=&0.
\end{eqnarray}

For massive particles, direct calculation shows $mv^a$ and $p^a$ satisfy the following relation\cite{PhysRevD.93.084025,Phys. Rev. D 58: 064005}:
\begin{eqnarray}
mv^a-p^a&=&\frac{S^{ab} R_{bcde} p^c S^{de}}{2(m^2+\frac{1}{4}R_{bcde} S^{bc} S^{de})}\label{vaua}.
\end{eqnarray}
The above formula shows that usually the velocity $v^a$ and the special momentum $u^a$ are not equivalent to each other. However, in the three-dimensional BTZ black hole spacetime, the right hand side of \eqref{vaua} vanished, i.e. $v^a=u^a$\cite{yuanzhang19}.
Moreover, if the spacetime possess some symmetries, then the corresponding Killing vector fields $\xi_a$ can be used to construct certain conserved quantities $Q_\xi$ of the particles as
\begin{eqnarray}
Q_\xi&=&p^a \xi_a+\frac{1}{2}S^{ab} \nabla_a \xi_b.
\end{eqnarray}

\subsection{Conserved quantities in the BTZ spacetime}
In this section, we consider  the BTZ black  hole spacetime and calculate the corresponding conserved quantities. The metric of the three-dimensional BTZ black hole reads\cite{Phys. Rev. Lett.1849}
\begin{eqnarray}
ds^2&=&-D({r})dt^2+D({r})^{-1}dr^2+r^2(-\frac{4GJ}{r^2}dt+d\varphi)^2,
\end{eqnarray}
where $D({r})=-8GM+\frac{r^2}{l^2}+\frac{16G^2J^2}{r^2}$ with $l=\sqrt{-\frac{1}{\Lambda}}$. $G$ is the constant of gravitation, $M$ and $J$ are the mass and angular momentum of the black hole respectively. If the gravitational mass and the angular momentum satisfy the condition $M\ge|J|/l$, the BTZ black hole will have an outer horizon and an inner horizon, which are located  at $r=r_+$ and $r=r_-$ respectively. In this paper, we consider the extremal BTZ black hole that $|J|=Ml$. Then, the outer and inner horizon coincides as $r_+=r_-=2l\sqrt{GM}$.

We choose an orthonormal basis for our calculations as follows
\begin{eqnarray}
e_a^{(0)}&=&\sqrt{D(r)}(dt)_a,\\
e_a^{(1)}&=&\frac{1}{\sqrt{D(r)}}(dr)_a,\\
e_a^{(2)}&=&r(d\varphi)_a-\frac{r_+ r_-}{lr}(dt)_a.
\end{eqnarray}
Note that the BTZ metric is stationary and axi-symmetric, we have two Killing vector fields: The first one is the timelike vector field $\xi_a$ and the second one is the axial Killing vector field $\Phi_a$. Under our orthonormal basis, they takes the form
\begin{eqnarray}
\xi_a&=&-\sqrt{D(r)}e_a^{(0)}-\frac{r_+r_-}{lr^2}e_a^{(2)},\\
\Phi_a&=&re_a^{(2)}.
\end{eqnarray}
By using the above two Killing vector fields, the particle's energy $E$ and total angular momentum $J$ can be obtained as
\begin{eqnarray}
E&=&-u^a\xi_a+\frac{1}{2m}S^{ab}\nabla_b\xi_a,\label{conservedE}\\
J&=&u^a\Phi_a-\frac{1}{2m}S^{ab}\nabla_b\Phi_a.\label{conservedJ}
\end{eqnarray}.
\subsection{Equations of motion on the equatorial plane}

Combining Eqs \eqref{equation one}, \eqref{equation two} and \eqref{equation four}, we can get the $(2+1)$-dimensional particle's antisymmetric spin tensor
\begin{eqnarray}
S^{(b)(c)}=\varepsilon^{(b)(c)}{}_{(d)}p^{(d)}s,
\end{eqnarray}
where $\varepsilon_{(b)(c)(d)}$ is a completely antisymmetric tensor and $\varepsilon_{(0)(1)(2)}=1$. Then the non-zero components of spin tensor is shown as
\begin{eqnarray}
S^{(0)(1)}=-sp^{(2)},\quad
S^{(0)(2)}=sp^{(1)},\quad
S^{(1)(2)}=sp^{(0)}.
\end{eqnarray}
Substitute the above equation into \eqref{conservedE} and \eqref{conservedJ}, the particle's energy and angular momentum read
\begin{eqnarray}
\label{energy of the particle}
E=\sqrt{D(r)}p^{(0)}+(\frac{r_+r_-}{lr}+\frac{rs}{l^2})p^{(2)},\\
\label{angular momentum of the particle}
J=s\sqrt{D(r)}p^{(0)}+(\frac{r_+r_-s}{lr}+r)p^{(2)},
\end{eqnarray}
which in turn gives
\begin{eqnarray}
p^{(0)}&=&\frac{l(El(lr^2+r_+r_-s)+J(lr_+r_-+r^2s))}{r\sqrt{(r^2-r_+^2)(r^2-r_-^2)}(l^2-s^2)},\\
p^{(2)}&=&l^2\frac{J-Es}{r(l^2-s^2)}.
\end{eqnarray}
In the coordinate basis, the particle's momentum can be expressed as
\begin{eqnarray}
\frac{dt}{d\tau}&=&\frac{1}{\sqrt{D(r)}}v^{(0)},\label{dtdtau}\\
\frac{dr}{d\tau}&=&\sqrt{D(r)}v^{(1)},\label{drdau}\\
\frac{d\varphi}{d\tau}&=&\frac{1}{r}v^{(2)}+\frac{r_+r_-}{lr^2\sqrt{D(r)}}v^{(0)}.\label{dphidtau}
\end{eqnarray}
The normalization condition of momentum is $p^a p_a=x$, where $x=-m^2$ for massive particles, or $x=0$ for massless particles respectively. In the following, in order to simplify the calculation, the mass of black hole is set to be unit, i.e. $M=1$. Then, according to the normalization condition, for massive particle we have
\begin{eqnarray}
\label{masslessnormalization}
u^{(1)}&=&\sigma\sqrt{(u^{(0)})^2-(u^{(2)})^2-1},
\end{eqnarray}
where $\sigma=-1$ corresponding the ingoing direction and $\sigma=1$ is the outward direction.
Collecting all the above ingredients, the momentum of the massive particles can be expressed as
\begin{eqnarray}
p_{1}^t(r)&=&\frac{B_{1} (r)}{D(r)},\label{timelikeconstraint}\\
\label{massiveparticlep1}
p_{1}^r(r)&=&\sigma\sqrt{C_{1} (r)},\\
p_{1}^\varphi(r)&=&\frac{r_+ r_-B_1(r)}{lr^2D(r)}+\frac{l^2(J-Es)}{mr^2(l^2-s^2)},
\end{eqnarray}
where
\begin{eqnarray}
B_1(r)=\frac{El(lr^2+r_+r_-s)-J(lr_+r_-+r^2s)}{mr^2(l^2-s^2)}\label{Br},
\end{eqnarray}
and
\begin{eqnarray}
C_1(r)=B_{1}^2(r)-\frac{1}{m^2}(m^2+(l^2\frac{J-Es}{r(l^2-s^2)})^2)D(r).
\end{eqnarray}
\subsection{Constraints on the orbits}
In the previous section, we already obtain the expression of the momentum for the massive particle. Note that there still exists some extra constraints such as the timelike constraint, which means not all possible orbits at this stage is permissible. Now we come to analyze  these constraints.

For the massive particle case, we introduce a critical angular momentum as follows\cite{yuanzhang19}
\begin{eqnarray}
J_c&=&\frac{lr_++r_-s}{r_+s+r_-l}El.
\end{eqnarray}
It is easy to see from Eq.\eqref{Br}, $B_{1}(r)=0$ when the angular momentum is equal to the critical value $J_c$. Note that the timelike constraint \eqref{timelikeconstraint} implies $B_{1}(r)>0$. Hence, for spinless particles($s=0$), the timelike constraint indicates $J \leqslant J_c$. While, for spinning particles, $J \leqslant J_c$ also implies $l^2-s^2>0$ which further constrain the massive particle's spin into the range of $-l<s<l$.

In collisional Penrose process, we consider the particles are falling from infinity and collides near the black hole horizon, and then one particle escape to infinity with more energy can either be massive or  massless\cite{Phys. Rev. D 86 024027,Maximal efficiency of the collisional Penrose process with spinning particles,Escape probability of the super-Penrose process,Maximal efficiency of the collisional Penrose process}. However, in extreme BTZ black bole, the escaped particle can only be massless, because
\begin{eqnarray}
 \lim_{r\to \infty }C_{1}(r) =-\infty <0,
\end{eqnarray}
which means the massive particle can not escape to the infinity\cite{yuanzhang19,Classical Quantum Gravity 11 2731}. Therefore, in this paper, we only consider the collision of one massive particle with another massless particle(PMP), which is usually referred as the Compton scatting in the literature\cite{Phys. Rev. D 86 024027,Maximal efficiency of the collisional Penrose process with spinning particles,Maximal efficiency of the collisional Penrose process}.

For massless particle, after combining Eqs. \eqref{energy of the particle}, \eqref{angular momentum of the particle}, the momentum of the particle can be expressed as
\begin{eqnarray}
p^{(0)}&=&\frac{Elr^2-Jr_+r_-}{r\sqrt{(r^2-r_{+}^2)(r^2-r_{-}^2)}},\\
p^{(2)}&=&\frac{J}{r}.
\end{eqnarray}
According to the normalization condition $p^a p_a=0$, we find
\begin{eqnarray}
\label{massivenormalization}
p^{(1)}&=&\sigma \sqrt{(p^{(0)})^2-(p^{(2)})^2}.
\end{eqnarray}
We assume $p^a=Ev^a$ by choosing a appropriately affine parameter $\rho$ \cite{Maximal efficiency of the collisional Penrose process with spinning particles}. Combining this fact with the momentum equations \eqref{dtdtau}-\eqref{dphidtau}, we can obtain
\begin{eqnarray}
p_{2}^{t}(r)&=&\frac{B_{2}(r)}{D(r)},\\
\label{masslessparticlep1}
p_{2}^{r}(r)&=&\sigma\sqrt{C_{2}(r)},\label{p2r}\\
p_{2}^{\varphi}(r)&=&\frac{r_+r_-B_{2}(r)}{D(r)r^2}+\frac{J}{r^2E},
\end{eqnarray}
where
\ba
B_{2}(r)&=&\frac{Elr^2-Jr_+r_-}{r^2lE},\\
C_{2}(r)&=&B_{2}^2(r)-\frac{J^2}{r^2E^2}D(r).
\ea
For extreme BTZ black hole, $r_+=r_-=2l$.
A massless particle can fly to infinity if and only if
\begin{eqnarray}
\lim_{r\to \infty }C_{2}(r) =E^2-\frac{J^2}{l^2}\geq0,\label{C2r}
\end{eqnarray}
which implies $-El\leq J\leq El$. Moreover, $J=El$ will lead to $p_{2}^{r}(r)$ being always equal to 0 and should be discarded. So the massless particle's angular momentum is constrained to $-El\leq J<El$.

\section{Collisional Penrose process of BTZ spacetime}

Note that the massive particle can not escape to infinity in BTZ spacetime, therefore in this section, for collisional Penrose process, we will only consider the possibility that one massless particle and another massive particle collide near an extreme BTZ black hole. After the collision, the massless particle has positive energy and escape to infinity. This process usually referred as Compton scattering \cite{Phys. Rev. D 86 024027,Maximal efficiency of the collisional Penrose process}. We are going to calculate the maximum energy extraction efficiency of this process. Moreover, for massive particles, we assume that their mass are all equal to $m$.

Note that during the collision process the momentum and spin of the whole system are conserved \cite{Phys. Rev. D 86 024027,Ultrahigh-Energy Debris from the Collisional Penrose Process,PhysRevD.97.064024}. That is
\begin{eqnarray}
p_{3}^{(\mu)}+p_{4}^{(\mu)}&=&p_{1}^{(\mu)}+p_{2}^{(\mu)},\\
S_{3}^{\mu \nu}+S_{4}^{\mu \nu}&=&S_{1}^{\mu \nu}+S_{2}^{\mu \nu}.
\end{eqnarray}
In the Compton scattering, the above equations could simplified as
\begin{eqnarray}
s_2&=&s_4,\\
\label{momentum conserved equation}
p_{3}^{(1)}+p_{4}^{(1)}&=&p_{1}^{(1)}+p_{2}^{(1)}.
\end{eqnarray}
Similarly, we can obtain the conservation of energy and total angular momentum:
\begin{eqnarray}
\label{energyconserved}
E_3+E_4&=&E_1+E_2,\\
\label{angularmomentumconserved}
J_3+J_4&=&J_1+J_2.
\end{eqnarray}

\subsection{The expansion of momentum near the horizon}
 In Compton scattering process, by employing Eqs. \eqref{massivenormalization} and \eqref{masslessnormalization}, the radial components of momenta for particle 1 (massless particle) and particle 2 (massive particle) can be written as
\begin{eqnarray}
\label{masslessnonspinp1}
p_{1}^{(1)}&=&\sigma_1\sqrt{-\frac{(J_1-E_{1}l)(E_{1}lr^2+J_1(-8l^2+r^2))}{-4l^2+r^2}},\\
\label{massivespinp1}
p_2^{(1)}&=&\sigma _2 \sqrt{\frac{l^2\left(J_2\left(4 l^3+r^2 s_2\right)-E_2\left(4 l^2 s_2+l r^2\right)\right){}^2}{r^2 \left(r^2-4l^2\right)^2 \left(l-s_2\right){}^2 \left(l+s_2\right){}^2}-\left(\frac{l^4 \left(J_2-E_2 s_2\right){}^2}{r^2 \left(l^2-s_2^2\right){}^2}+m^2\right)}.
\end{eqnarray}
For massive spinless particle, the Eq. \eqref{massivespinp1} reduces to
\begin{eqnarray}
\label{massivenonspinp1}
p_{2}^{(1)}&=&\sigma_2\sqrt{\frac{-8E_2J_2l^3+E_2^2l^2r^2+J_2^2(8l^2-r^2)-(-4l^2+r^2)^2}{(-4l^2+r^2)^2}}
\end{eqnarray}
Note that the particles collide near the horizon. The collision point $r_c$ is very close to the horizon, and therefore we set $r_c=2l/(1-\epsilon)$( $\epsilon\to0$ and $ \epsilon>0$  ). The  momentum $p_{\mu}^{(1)}$ can be expanded as
\begin{eqnarray}
p_{1}^{(1)}=\frac{|J_1-E_1l|\sigma_1}{4l\epsilon}+O(\epsilon^0),
\end{eqnarray}
and
\begin{eqnarray}
p_{2}^{(1)}=\frac{|J_2-E_2l|\sigma_2}{4(l-s)\epsilon}+O(\epsilon^0).
\end{eqnarray}
In addition, for the near-critical particle, the angular momentum reads $J=El+O(\epsilon)$, and similarly, for the noncritical particle, $J=El+O(\epsilon^0)$\cite{Maximal efficiency of the collisional Penrose process with spinning particles}. For outgoing massless particles, the critical value angular momentum $J=El$ is already excluded by Eq.\eqref{p2r}, therefore, here we first assume that the angular momentum of outgoing massless particle 3 is near-critical.

Then according to the momentum conserved equation \eqref{momentum conserved equation}, we can get
\begin{eqnarray}
\frac{|J_1-E_1 l|}{4l}\sigma_1+\frac{|J_2-E_2 l|}{4(l-s_2)}\sigma_2=\frac{|J_3-E_3 l|}{4l}\sigma_3+\frac{|J_4-E_4 l|}{4(l-s_4)}\sigma_4+O(\epsilon).
\end{eqnarray}
Similar to \cite{Maximal efficiency of the collisional Penrose process with spinning particles}, we find the case that particle 1 is near-critical while particle 2 is non-critical  will gives an attractive energy extraction efficiency and we also adapt this setup such that $\sigma_2=\sigma_4$ and $s_2=s_4$.

Since the massless particles is near critical, we have:
\begin{eqnarray}
\label{j1ande1}
J_1&=&E_{1}l(1-\lambda),\\
\label{j2ande2}
J_3&=&E_{3}l(1+\alpha_3 \epsilon+\beta_{3}\epsilon^2+...),
\end{eqnarray}
where $\lambda\to0^+$ and $\alpha_3$, $\beta_3$ are parameters of $O(\epsilon^0)$ with $\alpha_3<0$.

For particle 2, we assume that
\begin{eqnarray}
J_2&=&E_2l(1+\zeta),
\end{eqnarray}
the value range of $\zeta$ is $\zeta<0$ and $\zeta=O(\epsilon^0)$.

By substituting them into the Eqs. \eqref{energyconserved} and \eqref{angularmomentumconserved}, we can get
\begin{eqnarray}
\label{j4ande4}
J_4&=&E_{4}l(1+\frac{E_2}{E_4}\zeta+...).
\end{eqnarray}
Moreover, by calculating $E_2$ and $E_3$ separately. The energy extraction efficiency reads
\begin{eqnarray}
\eta=\frac{E_3}{E_1+E_2}.
\end{eqnarray}

\subsection{Maximal extraction efficiency for spinless particles}

To simplify calculation and without loss of generality, in the later calculation, we will set $m=1$. If the massive particle is spinless, plugging \eqref{j1ande1}-\eqref{j4ande4} into  \eqref{masslessnonspinp1} and \eqref{massivenonspinp1} respectively, and decomposing the momentum near the horizon ($r_c=\frac{2l}{1-\epsilon}$), we can get
\begin{eqnarray}
\label{momentum expand of particle 1}
p_1^{(1)}&=&\frac{E_1 \lambda }{4}\sigma_1\epsilon^{-1}+\frac{1}{8} E_1(4-5 \lambda )\sigma_1-\frac{E_1\left(\lambda ^2-8 \lambda +8\right) \epsilon }{16 \lambda }\sigma_1+O\left(\epsilon ^2\right),\\
\label{momentum expand of particle 2}
p_2^{(1)}&=&-\frac{E_2 \zeta }{4} \sigma_2\epsilon^{-1}+\frac{1}{8} E_2(5 \zeta +4)\sigma_2+\frac{\left(E_2^2\left(\zeta ^2+8 \zeta +8\right)+32\right) \epsilon }{16 E_2 \zeta }\sigma_2+O\left(\epsilon ^2\right),\\
\label{momentum expand of particle 3}
p_3^{(1)}&=&\frac{1}{4} E_3 \sqrt{\left(\alpha _3-4\right) \alpha _3}\sigma_3-\frac{  E_3\left(-2 \alpha _3\left(\beta _3+3\right)+5 \alpha _3^2+ 4 \beta_3 \right)\epsilon}{8 \sqrt{\left(\alpha _3-4\right) \alpha _3}}\sigma_3+O\left(\epsilon ^2\right),
\end{eqnarray}

\begin{small}
	\begin{eqnarray}
  \label{momentum expand of particle 4}	
	p_4^{(1)}&=&\frac{E_1 \lambda-E_2 \zeta}{4}\sigma_4\epsilon^{-1} +\frac{ 2 E_3\left(\alpha _3-2\right)+E_2(5 \zeta +4)+E_1(4-5 \lambda )}{8}\sigma_4 \notag\\
	&&+\frac{\epsilon}{16 (E_2 \zeta -E_1 \lambda )}\bigg(-2 E_1 \left(E_3 \lambda  \left(-5 \alpha _3+2 \beta _3-4\right)+E_2 \zeta  (\lambda -4)+4 E_2 (\lambda -2)+8 E_3\right) \notag\\
	&&-2 E_2 E_3 \left(\zeta  \left(5 \alpha _3-2 \beta _3+4\right)+8\right)+E_2^2 \left(\zeta ^2+8 \zeta +8\right)+E_1^2 \left(\lambda ^2-8 \lambda +8\right)+8 \left(E_3^2+4\right)\bigg)\sigma_4 \notag\\
	&&+O\left(\epsilon ^2\right).
	\end{eqnarray}
\end{small}

If both of the colliding particles are falling from infinity which means $\sigma_1=\sigma_2=-1$ and $\sigma_3=-\sigma_4=1$. However, we found that in this case all of the particles have no turning point. So in this paper, we consider that one of particles involved in the collision is produced by the previous scattering events in the ergoregion of the black hole\cite{Ultrahigh-Energy Debris from the Collisional Penrose Process}. Then the remaining possibility is that the massless particle is produced by the previous scattering events while the massive particle is falling from infinity such that $\sigma_2=-1$, and this in turn implies $\sigma_4=-1$ either. While $\sigma_1$ can be $-1$ or $+1$.

We found that if $\sigma_1=-1$, the calculation shows that $E_3=0$ which means no energy extracted from black hole. Therefore we set $\sigma_1=+1$ \cite{Maximum efficiency of the collisional Penrose process,Consistent analytic approach to the efficiency of collisional Penrose process}.
By employing the $\epsilon^0$ order of the momentum conservation equation \eqref{momentum conserved equation}, we obtain
\begin{eqnarray}
E_1-\frac{5E_1\lambda }{4}=\frac{1}{4}(\alpha _3-2-\sqrt{\left(\alpha _3-4\right) \alpha _3}) E_3 ,
\end{eqnarray}
or equally
\begin{eqnarray}
 E_3=S_1E_1,
 \end{eqnarray}
where
\begin{eqnarray}
S_1=\frac{ (5 \lambda -4)}{\alpha _3-2-\sqrt{\left(\alpha _3-4\right) \alpha _3}}.
\end{eqnarray}
Also, the $\epsilon^1$ order of the momentum conservation equation\eqref{momentum conserved equation} gives us
\begin{small}
	\begin{eqnarray}
    \label{equation of epsilon1}	
	&&\frac{1}{16}\bigg[\frac{2E_3 \left(-2 \left(\alpha _3 \left(\beta _3+3\right)\right)+5 \alpha _3^2+4 \beta _3\right)}{\sqrt{\left(\alpha _3-4\right) \alpha _3}}-\frac{E_2^2 \left(\zeta ^2+8 \zeta +8\right)+32}{E_2 \zeta }-\frac{E_1 \left(\lambda ^2-8 \lambda +8\right)}{\lambda } \notag\\
	&&+\frac{1}{E_2 \zeta -E_1 \lambda}\bigg(-2 E_1 \left(E_3 \lambda  \left(-5 \alpha _3+2 \beta _3-4\right)+E_2 \zeta  (\lambda -4)+4 E_2 (\lambda -2)+8 E_3\right) \notag\\
	&&-2 E_2 E_3 \left(\zeta  \left(5 \alpha _3-2 \beta _3+4\right)+8\right)+E_2^2 \left(\zeta ^2+8 \zeta +8\right)+E_1^2 \left(\lambda ^2-8 \lambda +8\right)+8 \left(E_3^2+4\right)\bigg)\bigg] \notag\\
	&&=0,
	\end{eqnarray}
\end{small}
when $\lambda \to 0$. It can be seen from Figure 1 that when $\alpha_3 \to 0$, $S_1$ takes the maximal value of 2. Since $\alpha_3$ is in the denominator of the decomposition term in \eqref{momentum expand of particle 3}, $\alpha_3$ can't be equal to zero. Numerical calculation is then carried out by substituting $\alpha_3$ approaching to zero.
\begin{figure}[!htb]
	\includegraphics [width=0.5\textwidth]{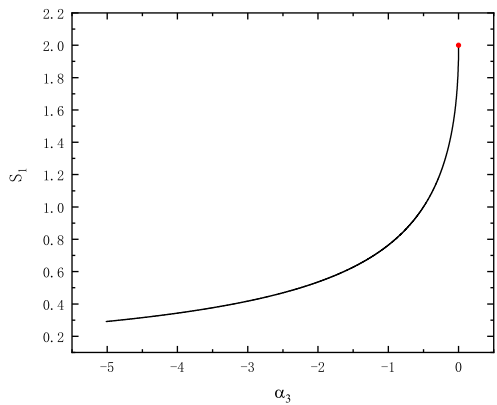}
	\caption{ The relationship between $S_1$ and $\alpha_3$ ($\lambda \to 0$). $S_1$ increases with the increase of $\alpha_3$. When $\alpha_3$ approaches to zero, the maximal value of $S_1$ is about 2.}
	\label{conditionofsFordiffrerentb}
\end{figure}
\begin{figure}[!htb]
	\includegraphics [width=0.5\textwidth]{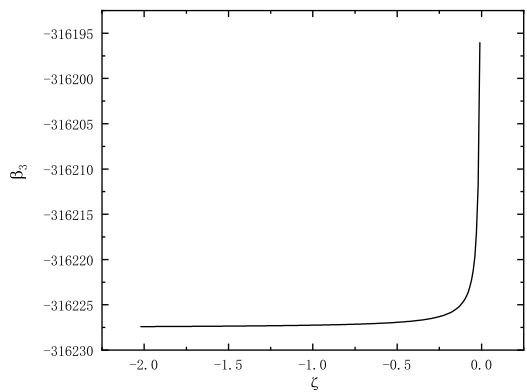}
	\caption{ The constraint of $\zeta$ and $\beta_3$ to realize large $E_1=1000$ when $E_2=1$ with $\lambda$ and $\alpha_3$ are chosen as $\lambda=0.000000001$ and $\alpha_3=0.0000001$.}
	\label{conditionofsFordiffrerentc}
\end{figure}

The energy of particle 1 and 2 should be $E_1>0$ and $E_2\geq 1$. Note that $E_3$ is proportional to $E_1$, if $E_1$ is much greater than $E_2$, then in the denominator we can safely ignore the small quantity $E_2$ by simply set $E_2=1$. From Fig. \ref{conditionofsFordiffrerentc}, we can see this is indeed the case, where we set $E_2=1$ and very large $E_1=1000$ can be realized with admissible values of $\zeta$ and $\beta_3$.

Collecting all the ingredients, the maximum energy extraction efficiency of Compton scattering reads
 \ba
 \eta=\frac{E_3}{E_1+E_2} \approx \frac{2E_1}{E_1+1} \approx 2.
 \ea

\subsection{Maximum extraction efficiency for spinning particles}

Now we turn to the spinning particles. After substituting  \eqref{j2ande2}, \eqref{j4ande4} into \eqref{massivespinp1}, momentum is expanded separately by $\epsilon$ as
\begin{eqnarray}
\label{momentum expand of particle two}
p_2^{(1)}=f_{21}\epsilon^{-1}+f_{22}+f_{23} \epsilon +O_2\left(\epsilon ^2\right),\\
\label{momentum expand of particle four}
p_4^{(1)}=f_{41}\epsilon^{-1}+f_{42}+f_{43} \epsilon +O_2\left(\epsilon ^2\right),
\end{eqnarray}
where the expression of $f_{21}$, $f_{22}$, $f_{23}$, $f_{41}$, $f_{42}$, $f_{43}$ can be found in the Appendix.

 By employing the $\epsilon^0$ order of the momentum conservation equation \eqref{momentum conserved equation}, we obtain
\begin{eqnarray}
E_3=S_2 E_1,
\end{eqnarray}
where the  $S_2(\lambda \to 0)$ reads

\begin{eqnarray}
S_2=-\frac{8 l^2-4 l s_2-4 s_2^2}{\left.2\left(-2 l^2+2 l s_2+\alpha _3 l^2+\alpha _3 l s_2+(s_2{}^2-l^2\right)\sqrt{\left(\alpha _3-4\right) \alpha _3}\right)}.
\end{eqnarray}

\begin{figure}[!htb]
	\includegraphics [width=0.6\textwidth]{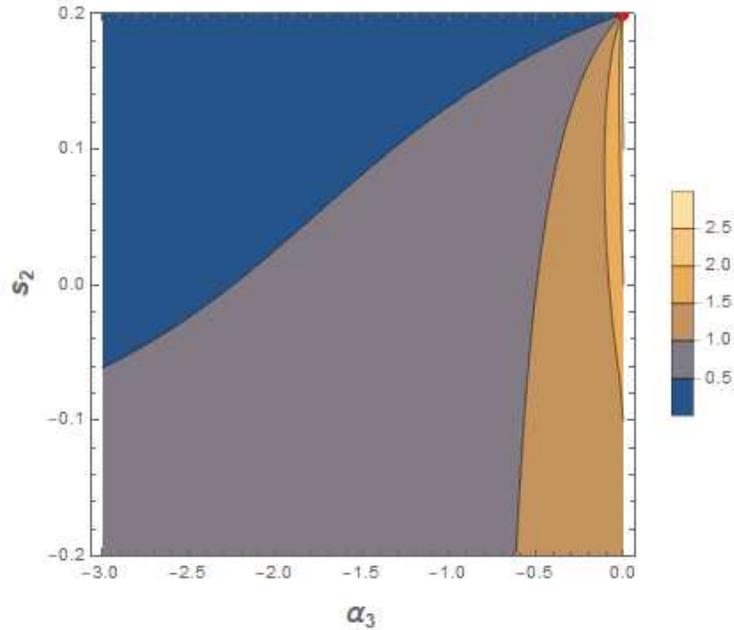}
	\caption{ The contour map of $S_2$ in terms of $s_2$ and $\alpha_3$ with $l=0.2$. The maximum value of $S_2$ is the red dot in the graph.}
	\label{conditionofsFordiffrerente}
\end{figure}

In order to obtain the maximum energy extraction efficiency $\eta$, we first calculate the maximum value of $S_2$. It can be seen from Fig. \ref{conditionofsFordiffrerente} that the red point in the graph $\left(\alpha _3,s_2\right)\approx (0,l)$, corresponds the maximum value of $S_2$. This value is approach 3 as the spin $s_2$ approach to $l$.

Since the maximum value of $E_3=S_2 E_1\approx3E_1$, for simplicity, we set the massive particle's energy $E_2=1$. Similar to the massless case, $E_1\gg E_2$ corresponds the highest extraction efficiency, that is
\ba
\eta_{max}=\frac{E_3}{E_1+E_2}\approx \frac{3 E_1}{E_1+1}\approx 3.
\ea We can verify that for $\left(\alpha _3,s_2\right)\approx (0,l)$, given a value of $E_1$, we can always find the $\zeta$ and $\beta_3$ satisfying the constraint conditions makes $E_2$ being the minimum value.

Moreover, our calculation shows that the maximum energy extraction efficiency $\eta_{max}$ of BTZ black hole is independent with the value of the cosmological constant. The reason for that is because when $\alpha_3 \to 0$, the maximum of $S_2$ can only be achieved as $s/l$ approach to $1$. Therefore the maximum value of $S_2$ is independent with value of $l$.

\section{CONCLUSIONS}
In this paper, we study the collisional Penrose process in the vicinity of an extremal BTZ black hole. Due to the existence of negative cosmological constant, we found that only massless particles could escape to infinity. Hence, the admissible collisional Penrose process must happened between one massless particle collides with a massive particle near the horizon of BTZ black holes. After collision, the massless particle escape to infinity with energy larger than the incident particles.

By employing the MPD equations which describing the motion of the particles, we explicitly obtained the momentum and the velocity of the particles. Moreover, by applying timelike constraint, we found a restriction on the angular momentum of the particles, namely, the angular momentum of the particle can not exceed the critical value $J<J_c=El$.

With the help of the conservation of energy and momentum during the collision, we calculate the maximum energy extraction efficiency $\eta_{max}$ of the collisional Penrose process, and the detailed calculation is performed for both spinless and spinning particles. Our results show that the spinning particle have a higher energy extraction efficiency than the spinless particle. More specifically, for spinless ingoing massive particle, the maximum energy extraction efficiency $\eta_{max}\approx 2$. While for the spinning massive particle, the maximal value is raised to $\eta_{max}\approx 3$.

Last but not least, our calculation also indicates that the maximum energy extraction efficiency is independent of the value of the cosmological constant of BTZ black holes. This conclusion is valid no matter the massive ingoing particle has spin or not.

\begin{acknowledgements}

This work is supported by NSFC with No.11775082. The authors would like to thank Xulong Yuan and Yunlong Liu for helpful discussions.

\end{acknowledgements}

\section{APPENDIX}
\begin{small}
	\begin{eqnarray}
f_{21}&=&-\frac{E_2 \zeta  l}{4 l-4 s_2}\sigma_2\\
f_{22}&=&\frac{E_2 l\left(l(5 \zeta +4)+s_2(\zeta -4)\right)}{8 \left(l-s_2\right) \left(l+s_2\right)}\sigma_2  \\
f_{23}&=&\frac{1}{16 E_2 \zeta l \left(l-s_2\right) \left(l+s_2\right){}^2}[\left(E_2^2 \left(\zeta ^2+8 \zeta +8\right)+32\right) l^4-2 E_2^2 l^3 \left(\left(3 \zeta ^2+8 \zeta +8\right) s_2\right)               \notag\\
&&+l^2 \left(\left(E_2^2 \left(\zeta ^2+8 \zeta +8\right)-64\right) s_2^2\right)+32 s_2^4]\sigma_2
     \end{eqnarray}
\end{small}

\begin{eqnarray}
f_{41}&=&-\frac{(E_2 \zeta -E_1 \lambda)l}{4 l-4 s_2}\sigma_4  \\
f_{42}&=&\frac{l \left(2 E_3 l\left(\alpha _3-2\right) +2 E_3 s_2\left(\alpha _3+2\right) +E_2 (\zeta -4) s_2\right)}{8 \left(l-s_2\right) \left(l+s_2\right)} \notag\\
&&-\frac{l \left(E_2 (5 \zeta +4) l-E_1 \left((5 \lambda -4) l+(\lambda +4) s_2\right)\right)}{8 \left(l-s_2\right) \left(l+s_2\right)}\sigma_4
\end{eqnarray}

\begin{eqnarray}
f_{43}&=&\frac{\Delta}{\Sigma}\sigma_4  \\
\Sigma&=&\frac{1}{16  l \left(l-s_2\right) \left(l+s_2\right){}^2(E_2 \zeta -E_1 \lambda)}\\
\Delta&=&32 s_2^4+l^2s_2^2\left(-2 E_3 E_2 \left(\zeta  \left(\alpha _3-2 \beta _3+4\right)+8\right)+E_2^2 \left(\zeta ^2+8 \zeta +8\right)+8 \left(E_3^2-8\right)\right. \notag\\
&&\left.E_1^2 \left(\lambda ^2-8 \lambda +8\right)-2 E_1 \left(-E_3 \lambda  \left(\alpha _3-2 \beta _3+4\right)+E_2 \zeta  (\lambda -4)+4 E_2 (\lambda -2)+8 E_3\right)\right)  \notag\\
&&+l^4\left(-2 E_3 E_2 \left(\zeta  \left(5 \alpha _3-2 \beta _3+4\right)+8\right)+E_2^2 \left(\zeta ^2+8 \zeta +8\right)+8 \left(E_3^2+4\right)\right. \notag\\
&&\left.E_1^2 \left(\lambda ^2-8 \lambda +8\right)-2 E_1 \left(E_3 \lambda  \left(-5 \alpha _3+2 \beta _3-4\right)+E_2 \zeta  (\lambda -4)+4 E_2 (\lambda -2)+8 E_3\right)\right)   \notag\\
&&-2l^3s_2\left(2 E_2 E_3 \left(3 \alpha _3 \zeta -2 \left(\beta _3+2\right) \zeta -8\right)+E_2^2 \left(3 \zeta ^2+8 \zeta +8\right)+E_1^2 \left(3 \lambda ^2-8 \lambda +8\right)+8 E_3^2\right. \notag\\
&&\left.-2 E_1 \left(E_3 \left(3 \alpha _3 \lambda -2 \left(\beta _3+2\right) \lambda +8\right)+E_2 (3 \zeta  \lambda -4 \zeta +4 \lambda -8)\right)\right)
\end{eqnarray}

\bibliographystyle{unsrt}

\end{document}